\documentclass[reprint,aps,pre,showpacs,twocolumn]{revtex4-2}

\usepackage{amsmath, amssymb}  
\usepackage{amsfonts} 
\usepackage{graphicx}
\usepackage{mathptmx, bm}
\usepackage{float}

\begin{document}

\title{Least Squares as Random Walks}

\author{Alexander Kostinski$^{1}$}
\email{kostinsk@mtu.edu}

\author{Glenn Ierley$^{2}$}
\email{Deceased}

\author{Sarah Kostinski$^{3}$}
\email{sk10775@nyu.edu}

\affiliation{\noindent \textit{$^{1}$Department of Physics, Michigan Technological University, Houghton, MI 49931 USA}} 

\affiliation{\noindent \textit{$^{2}$Emeritus; Scripps Institution of Oceanography, UC San Diego, La Jolla, CA 92093 USA}}

\affiliation{\noindent \textit{$^{3}$Department of Physics, New York University, New York, NY 10003 USA}}

\begin{abstract} 
\noindent

\noindent 

Linear least squares (LLS) is perhaps the most common method of data analysis, dating back to Legendre, Gauss and Laplace. Framed as linear regression, LLS is also a backbone of mathematical statistics. Here we report on an unexpected new connection between LLS and random walks. To that end, we introduce the notion of a random walk based on a discrete sequence of data samples (\textit{data walk}). We show that the slope of a straight line which annuls the net area under a residual data walk equals the one found by LLS. For equidistant data samples this result is exact and holds for an arbitrary distribution of steps.

\end{abstract}

\maketitle

Linear least squares (LLS) is arguably the oldest and most commonly used method of curve fitting and data analysis.  It dates back to Gauss' finding of the ``missing planet'' (asteroid Ceres), which was likely preceded by Legendre and further advanced by Laplace~\cite{stigler1981gauss}.  Later interpreted as linear regression by Galton and Pierson, LLS began to form a basis of modern mathematical statistics and data science \cite{Bevington92,stigler2016seven,papoulis1990probability}.  The term ``regression'' originated with Francis Galton, the cousin of Charles Darwin, and with Karl Pearson in the context of heredity~\cite{bulmer2003francis,porter2010karl}.  

LLS is fundamental in the theory of measurements and arises in all applied sciences. Three interpretations of LLS data fitting can be given \cite{papoulis1990probability}: (i) both $y$-values and $x$-values are deterministic; (ii) because of noise $y$-values are random variables but $x$-values are deterministic ($x$ being the ``control variable'' in experiments); (iii) both $y$- and $x$-values are random, e.g., arising in problems of prediction. The resulting formulae are identical for all of the interpretations~\cite[p.403]{papoulis1990probability}.  For the sake of clarity, we shall use case (ii) here as it is most commonly encountered in the physical sciences.  For broader context, we recall that when the fluctuations (``errors'') are normally distributed, the maximum likelihood principle yields the LLS estimate and more generally, the LLS estimate is a minimum variance one~\cite{papoulis1990probability,press2007numerical}.  One would think that given more than two centuries of research, nothing new can be said about LLS.  Nevertheless, in this letter we report on one such new and unexpectedly simple result by connecting LLS to random walks.  

To that end, we introduce the notion of a random walk based on data, or simply a ``data walk''. Although our result does not depend on the noise distribution nor on its mean, for the sake of clarity we illustrate the result on a set of scattered data points generated here by Mathematica \cite{wolfram1991mathematica} with arbitrary units and simplest default options: a linear function of unit slope, embedded in additive white zero-mean unit variance Gaussian noise denoted by ${\cal N}(0,1)$. The $x$-values are uniformly spaced on a unit interval and given by

\begin{figure*}
    \centering
    \includegraphics[width=\textwidth]{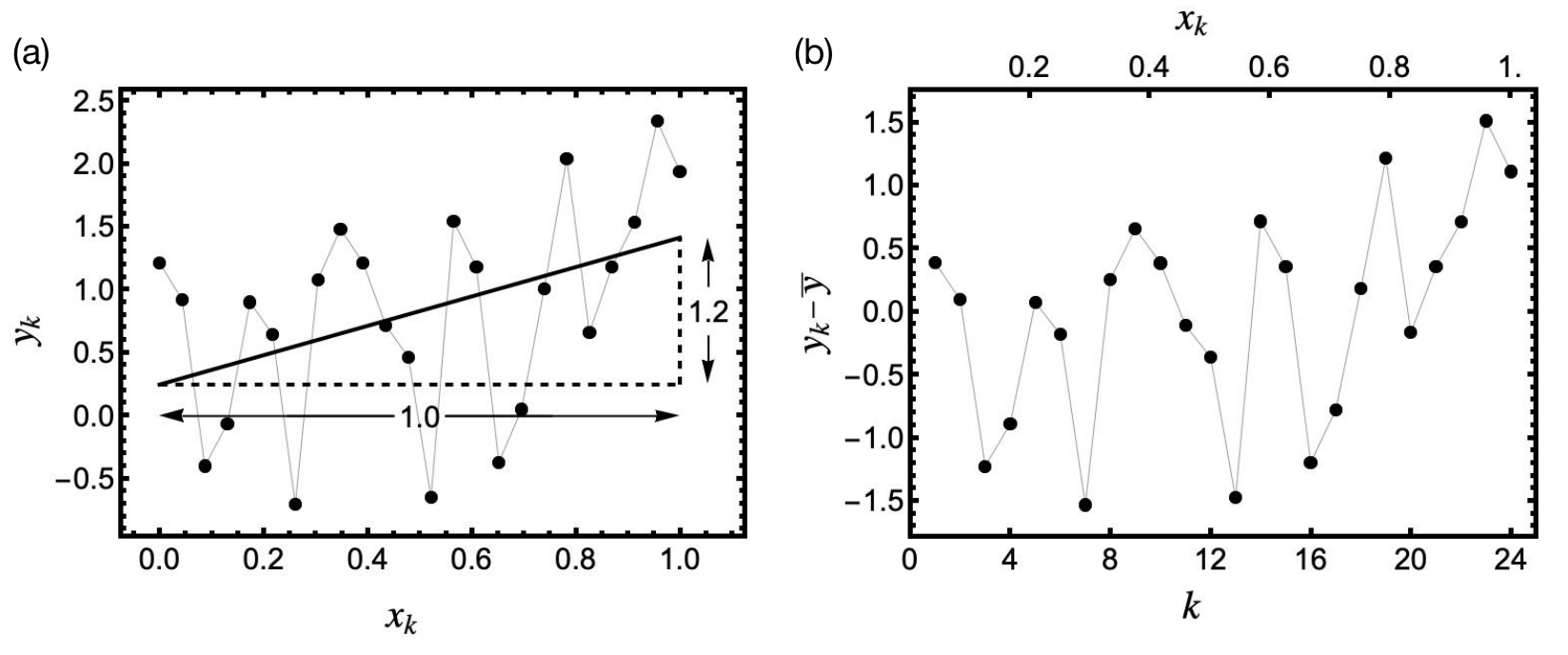}
    \vspace{-9mm}
    \caption{{\bf Linear function with additive Gaussian noise}. \emph{Panel (a)}: Twenty four data samples generated via $y_k = x_k + n_k$ vs.\ $x_k$ for $k=1,\ldots, 24$. The $n_k$s are independent random samples drawn from a zero-mean unit variance Gaussian probability distribution denoted by ${\cal N}(0,1)$. The solid  black line is a linear least squares (LLS) fit to this noisy data. The LLS slope estimate of $1.2$ differs from unity because of the sampling variability. \emph{Panel (b)}: same data samples as in (a) but shifted vertically because the arithmetic mean is subtracted from each sample.  The upper abscissa gives the original values while the lower one represents the transition to consecutive integers, subsequently interpreted as steps of a data walk (see text).}
    \label{fig:rawdata} 
\end{figure*}

\begin{equation}
x_k = \frac{k-1}{N-1} \qquad k = 1, 2, \ldots, N\, 
\label{eq:xk}
\end{equation}
and the corresponding y-values are
\begin{equation}
y_k = x_k + n_k
\label{eq:yk}
\end{equation}
for $k=1,2,\ldots,24$ ($N=24)$, with the resulting synthetic data shown in panel (a) of Fig.~\ref{fig:rawdata}.  The dataset used here, as well as the Mathematica code used to generate and plot the data, are provided as Supplementary Material~\cite{SM}.

Towards defining the key notion of a data walk (DW), we keep the ordinate values of Fig.~\ref{fig:rawdata}(a) but, in order to interpret the $x$-axis as history of a random walk, we replace the abscissa values by a sequence of consecutive integers, $k=1,2,\ldots,N$.  By doing so we discard information on dimensions and magnitude of the basic unit. However, because the $x$-values are evenly spaced, no other information is lost and the unit's magnitude and dimensions can always be restored once data fitting is completed. Note that this minimal loss of information does not hold if the data are not uniformly sampled, e.g., contain gaps. 

The transition is illustrated by panel (b) of Fig.~\ref{fig:rawdata} with the upper and lower $x$-axes. The consecutive integers of the abscissa are now regarded as steps of the DW, to be defined next. To that end, the arithmetic (sample) mean $\bar{y} \equiv \sum_{k=1}^N y_k / N$ is removed (ordinates shifted) in panel (b) of Fig.~\ref{fig:rawdata} and we let the data value excess for each $y_k$, that is, $y_k - \bar{y}$ shown in panel (b) of Fig.~\ref{fig:rawdata}, serve as successive steps (increments) of DW.  Thus the DW is a cumulative (running) sum defined by
\begin{equation}
z_j  = \sum_{k=1}^j \, (y_k - \bar{y})\qquad j=1, 2, \ldots, N
\label{eq:PRW}
\end{equation}
where we assign (prepend) $z_0 = 0$, pinning the DW at both ends since $z_N = 0$ because of the $\bar{y}$ subtraction in Eq.~(\ref{eq:PRW}).  

\begin{figure*}
    \centering
    \includegraphics[width=\textwidth]{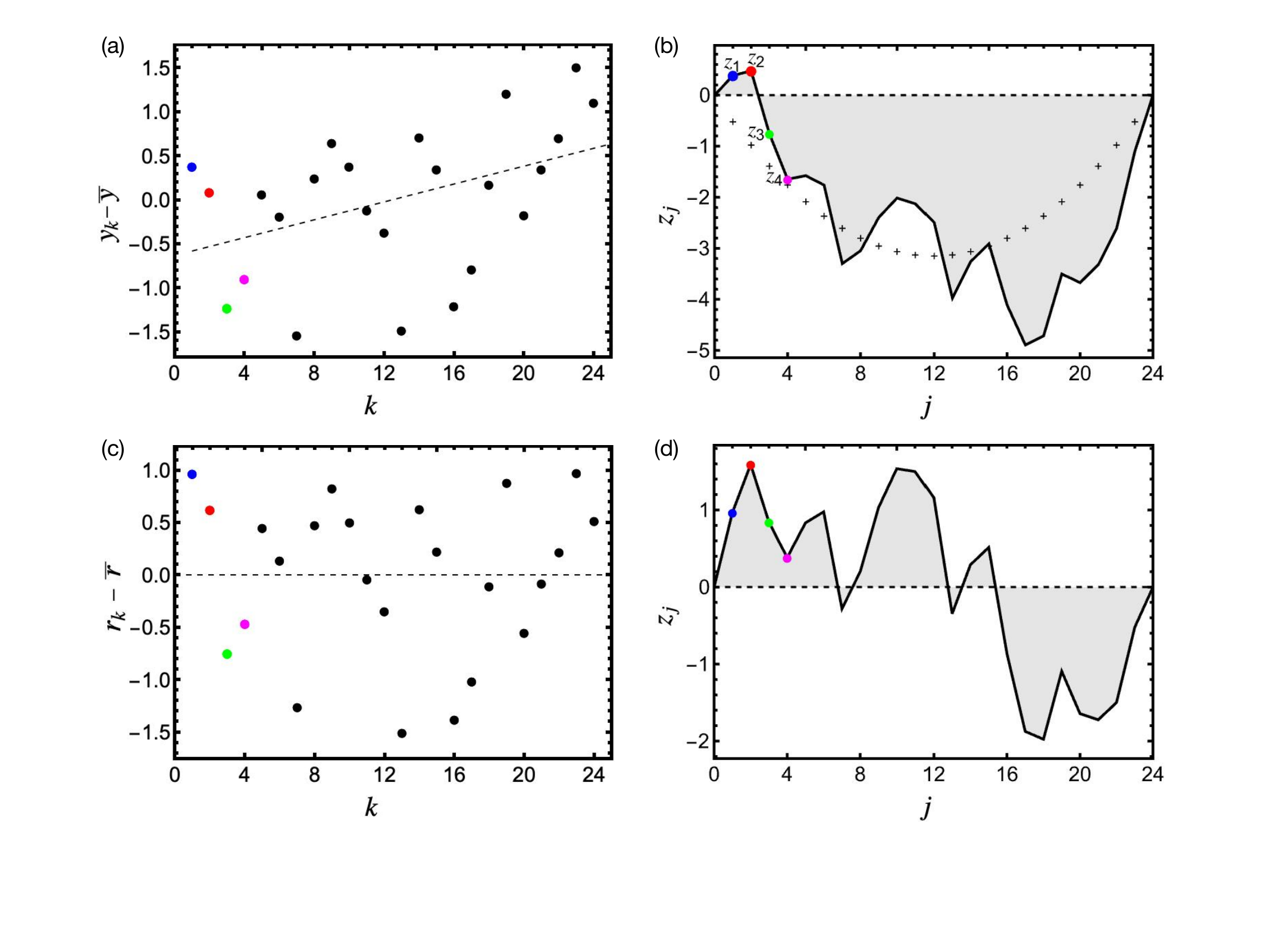}
    \vspace{-7mm}
    \caption{{\bf Construction of a pinned data walk (DW)}. \emph{Panel (a)}: Data string $y_k - \bar{y}$ from panel (b) of Fig.~\ref{fig:rawdata}, vs. step number $k=1, \ldots, 24$. The ordinates serve as increments (steps) to construct the DW via Eq.\ (\ref{eq:PRW}). There is an upward trend in the data sequence. \emph{Panel (b)}: The resulting DW has one interior zero, and an area under the curve of $58.3$ via Eq.\ (\ref{eq:trend}), indicating a large positive trend. The unit-slope trend, noise-free reference parabola (area of 50) given by Eq.~(\ref{eq:A5}) is illustrated with crosses.  \emph{Panel (c)}: Upon subtraction of the LLS fit from the data in panel (a), the residuals $r_k = y_k - \alpha x_k$, shifted by their sample mean $\bar{r}$, yield the DW bridge in panel (d). \emph{Panel (d)}: DW has five interior zeros, causing a perfect cancellation of signed areas and zero trend via Eq.\ (\ref{eq:trend}).  For a symmetric random walk (not necessarily a bridge), the mean number of zero-crossings is $\sim \! \!\sqrt{N}$ (here $5 \approx \sqrt{24}$) while the most likely number of zero-crossings is zero \cite{papoulis2002random}. The conclusion is that \textit{the LLS slope annuls the area under the residual DW and vice versa}, as proven in the main text. This theorem holds for an arbitrary (e.g., asymmetric) distribution of DW steps.}
    \label{fig:walks} 
\end{figure*}
Insofar as the data are corrupted by random errors and noise, these DW bridges described by the $z_j$s in Eq.~(\ref{eq:PRW}) are also random, albeit correlated by trends in the data. The $z_j$s based on data of Fig.~\ref{fig:rawdata} are depicted in panel (b) of Fig.~\ref{fig:walks} and can be regarded as successive positions of the data walker. DW is a bridge because return is enforced by subtraction of the sample mean. Owing to $\bar{y}$ subtraction in (\ref{eq:PRW}), the DW is independent of the intercept and we pose the question of finding the slope via DW. The finding is astonishingly simple, yet new to the best of our knowledge: the slope that annuls the (signed) area under the DW equals the LLS slope as shown in Fig.~\ref{fig:walks} and discussed next.

To gain intuition for the DW construction, let us briefly consider two extreme cases: (i) no signal (noise only), corresponding to the classical random walk; (ii) no noise (deterministic case). In the absence of signal, $y_k = n_k$ and the resulting $z_j  = \sum_{k=1}^j \, (n_k - \bar{n}); j=1, 2,\ldots N$, with $z_j$ representing the position of a symmetric random walker after $j$ steps.  Here the ``true'' mean is zero because we happen to know that the generated noise is zero-mean in the infinite $N$ limit. This standard random walk is symmetric and the traditional Gaussian-distributed length of the steps was chosen for Fig.~\ref{fig:rawdata}. In the absence of noise, the process is fully deterministic and the data are given by a straight line with slope $\alpha$: $y_k = \alpha \,x_k $.  There is no sampling variability and, therefore, no scatter of data points.  The intermediate case typically occurring in practice involves a signal embedded in some noise and it is the signal that renders the DW asymmetric, with the bias related to the signal-to-noise ratio, the latter being order unity for the data in Fig.~\ref{fig:rawdata}.  So now we ``let your data do the walking'' and state the main result.

To that end, we turn to Fig.~\ref{fig:walks} with a dual purpose: (i) to illustrate the DW construction geometrically and (ii) to observe the relation between the slope of the data and zero-crossing frequency of the DW. We color-coded the first few points to help the reader see the DW construction at a glance. For example, the blue point in Fig.\ \ref{fig:walks}(a) is positive, causing the ``walker'' executing the DW in Fig.\ \ref{fig:walks}(b) to move up from the origin.  This continues on both accounts for the red data point. The pattern reverses for subsequent green and purple points whose contributions are both negative and so on. Observe the contrast between panels (b) and (d) of Fig.~\ref{fig:walks}, for data with and without the trend, respectively. The association between the trends in the data and the frequency of their DW zero-crossings is evident. 

We digress briefly to summarize the DW construction in a friendlier continuous variable notation: $z(t) \equiv \int_0^t (y(x) - {\bar y}) \, dx$. Then, the measure of trend in the data is given by the area under the DW (see Eq.~\ref{eq:trend} below): $A(y) = \int_0^1 d\eta \int_0^\eta (y(x) - {\bar y}) \, dx$.  For the simplest reference case of $y(x) = x$ (no noise, unit slope), the DW is an upward parabola shifted down by the $\bar y$ subtraction: $\int (y(x)-{\bar y}) \, dx = x^2/2 - x/2$.  This parabola is paralleled by the discrete one given in Eq.~(\ref{eq:A5}) and shown by the crosses in panel (b) of Fig.~\ref{fig:walks}.

Signal presence typically causes largest values of \textit{accumulated deviations from the sample mean} near the middle of the DW bridge.  Conversely, in the absence of signal, one now expects noise alone to send the data walker up or down with equal likelihood. Hence, detrending might then be associated with symmetry and thus DW area annulment.

Numerical experiments whose constructions resemble those of Fig. \ref{fig:walks} led us to the following theorem: \\

{\em The slope of the fit to data that annuls the area under a DW equals that of the LLS fit}.  \\

\noindent The core of this letter ends here, and the remainder is devoted to proving this theorem, interpreting statistical significance of the results, and a few remarks.  

As just discovered, the area under the DW is a measure of a trend and we work with the negative of the net (signed) area under the DW, denoted by $A(y)$:
\begin{equation}
A(y)  =  -\sum_{j=1}^N\, z_j \, .
\label{eq:trend}
\end{equation}
Whereas $y_k$s are deviations from the sample mean, $z_j$s are accumulated deviations up to the current step (the data walker positions after $j$ steps). Then $A(y)$ is the relative amount of time that our data walker spends above the origin. The stronger the signal, the more time is spent in the upper (lower) half before the eventual return.  

Next, we work out the area-slope relation for the simplest case of a linear function in the absence of noise with uniformly sampled values at steps given by Eq.~(\ref{eq:xk}). We denote this noiseless unit slope reference area $A(y=x) \equiv A(x)$. Using Eqs.~(\ref{eq:xk})--(\ref{eq:trend}), a line with slope $\alpha$ in $y_k = \alpha \,x_k $  yields a DW net area:
\begin{equation}
A(y) = A(\alpha \, x) = \alpha \, A(x)  = \alpha \, \frac{N\, (N+1)}{12}\, , 
\label{eq:lintrend}
\end{equation}
To arrive at the rightmost expression of Eq.~(\ref{eq:lintrend}) (see \ref{appendixA} for a detailed derivation), we used the partial sum results:
\begin{equation}
    \sum_{k=1}^N \, k = \frac{N\, (N+1)}{2}
   \qquad
   \sum_{k=1}^N \, k^2 = \frac{N\, (N+1)\, (2N+1)}{6}\, ,
   \label{eq:triangularnumber}
\end{equation}
where the former sum is also known as a triangular number. For $N=24$, $N(N+1)/12 = 50$ but for our {\em noisy} data,  $A(y)=58.35$ and the slope $\alpha = 12/(24 \times 25) \times 58.35 = 58.3/50 = 1.17$ annuls the (signed) area of the residuals as shown in panel (d) of Fig.~\ref{fig:walks} and explained next. 

The areas defined in Eq.~(\ref{eq:trend}) are additive so that $A(y^{(1)} + y^{(2)}) = A(y^{(1)}) + A(y^{(2)})$. For the general noisy case, the area under the DW before detrending is $A(y)$ which, in contrast to $A(x)$, is a random variable, and depends on a sample realization of noise.  To annul (detrend) it, we set $\alpha\, A(x)  =  A(y)$ to obtain
\begin{equation}
    \alpha =  \frac{A(y)}{A(x)} = 
    - \frac{12}{N\, (N+1)}\, \sum_{k=1}^N\, z_k = 
    -\frac{12}{N\, (N+1)}\, \sum_{k=1}^N\, \sum_{j=1}^k (y_j - \bar{y}) \, .
    \label{eq:slope_tr}
\end{equation}
The underlying signal obscured by noise need not be a linear one and the function could be, say, an exponential or power-law rise (reducible to a linear one by a change of variable, or on semi-log, log-log plots, etc). More generally, this approach simply supplies a one-parameter means to quantify the trend. 

To relate the DW to LLS, recall that the conventional least squares slope is given by~\cite{Bevington92}
\begin{equation}
\alpha = \frac{N \sum x_k \, y_k - \sum x_k \sum y_k}{N \sum x_k^2 - (\sum x_k)^2}\, ,
\label{eq:LLS}
\end{equation}
arising from minimizing the goodness-of-fit metric $\chi^2$ (this sum of squared residuals is 20.6 for the data in Fig.~\ref{fig:rawdata} and the LLS fit of $y = 1.17 x + 0.24$).  For the equally spaced $x_k$s as in Eq.~(\ref{eq:xk}), $\sum_{k=1}^{N} x_k = N/2$ and $\sum_{k=1}^{N} x_k^2 =N(2N-1)/(6(N-1))$, so that Eq.~(\ref{eq:LLS}) reduces to
\begin{equation}
\alpha = \frac{12\, (N-1)}{N\, (N+1)}\, \sum_{k=1}^N \, y_k \left( x_k - \frac{1}{2} \right)\, .
\label{eq:simpleSlope}
\end{equation}

The seemingly unrelated random walk expression (\ref{eq:slope_tr}) for $\alpha$, found by annulling the area under the DW is, in fact, identical to (\ref{eq:simpleSlope}) as we now prove. One must show that
\begin{equation}
\sum_{k=1}^N\, \sum_{j=1}^k \left ( y_j - \bar{y} \right )  =  (N-1)\, \sum_{k=1}^{N} y_k \,\left  ( \frac{1}{2}  - \frac{k-1}{N-1} \right )\, .
\label{eq:lsq}
\end{equation}
By induction, the sum $\sum_{k=1}^{N} \sum_{j=1}^{k} y_j$ on the left gives $\sum_{k=1}^N \, (N-k+1)\, y_k\,$ (details provided in~\ref{appendixB}).  Meanwhile the second term in that sum is the constant $\bar{y}$, so that evaluation of the outer two sums gives
\begin{align}
 \sum_{k=1}^{N} \sum_{j=1}^{k} (-\bar{y}) 
 &= - \frac{N(N+1)}{2} \bar{y} = -\frac{N(N+1)}{2} \left(\frac{1}{N} \sum_{k=1}^{N} y_k \right)  \nonumber \\
 & = - \sum_{k=1}^N \, \frac{N+1}{2}\, y_k\, .
 \label{eq:meanysum}
\end{align}
When merged, these last two results yield
\begin{equation}
    \sum_{k=1}^N\, \sum_{j=1}^k \left ( y_j - \bar{y} \right ) = \sum_{k=1}^N \, y_k \left( \frac{N+1}{2} - k \right) 
    \label{eq:reduce}
\end{equation}
and this is the same as the right-hand side of (\ref{eq:lsq}) after distributing through the leading factor of $N-1$.  The equality of slopes from Equations (\ref{eq:slope_tr}) and (\ref{eq:simpleSlope}) is remarkable because the two approaches to defining a slope are so unlike each other: the LLS derivation has nothing to do with random walks and gives a two parameter fit (slope and intercept), while the DW route is independent of a constant offset.  However, we note that the DW approach can also be used to obtain the intercept: $\bar{r} = \frac{1}{N} \sum_{k=1}^{N} r_k = \overline{y - \alpha x}$ equals the LLS intercept.

We now address the statistical significance of the DW slope. To that end, we rewrite Eq.~(\ref{eq:reduce}) compactly by defining a zero-mean $N$-dimensional vector $\tilde{\bf x}$ with evenly spaced elements $\tilde x_k = ((N+1)/2) - k $ for $ k=1,2, \ldots N$.  Then, together with Eq.~(\ref{eq:slope_tr}), $A(y) = - {\bf y}\cdot\tilde{\bf x}$ where the $k$th element of the data vector ${\bf y}$ is $y_k$. As the $y_k$s are independent and identically distributed (IID) random variables with variance $\sigma^2$, the probability density function of $A(y)$ has variance
\begin{equation}
\sigma_A ^2 = \sigma^2\, \tilde{\bf{x}}\cdot\tilde{\bf{x}} 
 = \sigma^2 \, \sum_{k=1}^N\, \left (\frac{N+1}{2}-k \right )^2 =  \frac{(N^3 - N)}{12} \sigma^2 \
\label{eq:stdevSlope}
\end{equation}
where we have again used the results of Eq.~(\ref{eq:triangularnumber}).  For $N \gg 1$, the probability density function of $A(y)$ approaches a normal distribution by the central limit theorem.  Using Eq.~(\ref{eq:lintrend}) for $A(x)$, the DW slope of $A(y)/A(x)$ then has a variance of 
\begin{equation}
\frac{12(N-1)}{N\, (N+1)}\, \sigma^2 \, ,
\end{equation}
which is identical to the standard result~\cite{Bevington92} for LLS.  In passing, we note that the DW area-annulling slope 
$\alpha$ can be expressed in simple geometric terms as a projection of the data vector $\bf y$ onto $\tilde {\bf x}$, that is, $\alpha = (\bf{y} \cdot \tilde{\bf x})$ / $(\tilde{\bf x} \cdot \tilde{\bf x})$.

A conventional measure of significance in data analysis is deviation scaled by the standard error $\sigma_{SE}$: the so-called $t$-statistic~\cite{press2007numerical}.  The exact relation of our two $t$-statistics is 
\begin{equation}
t_{A} = \sqrt{\frac{N-1}{N-2}} \, \, t_{LS}
\label{eq:tstat}
\end{equation}
because the LLS standard error is for $N-2$ degrees of freedom, accounting for both a slope and an intercept, while the DW route estimates only the slope and hence $N-1$ appears in the numerator. Thus, not only are the DW and LLS slopes and variances equal, but their measures of significance are also consistent. 

As a specific illustration, consider the area (trend) for our data set of 24 points ($N = 24$) in Fig.~2(b), where we compute $A(y) = 58.35$. This is to be compared with the value of $\sigma_A = 28.4$ found from Eq.~(\ref{eq:stdevSlope}) for $N=24$ and a computed sample standard deviation of $\sigma=0.84$. This value of $A(y)$ hence represents a departure of $2.05\, \sigma_A$. If that ratio exceeds a user-chosen confidence level, the slope of $1.17$ calculated from Eq. (\ref{eq:simpleSlope}), is deemed significant. 

Thus $\sigma_A$ serves as a benchmark for a typical net area from $N$ noisy data points. For $N\gg1$, $\sigma_A \sim \sigma \, N^{3/2} / \sqrt{12}$. The large $N$ limit for area in the presence of a linear function is, from Eq.~(\ref{eq:lintrend}), $\alpha\, N^2/12$. For a given pair of $\sigma$ and $\alpha$, the area contribution for the signal grows as $N^2$ while the noise contribution growth is slower: $\sim \! N^{3/2}$. Inevitably, the signal emerges more accurately with increasing $N$, but only as $\sqrt{N}$. 

We now compare the above value of $t_A = 2.05$ to the calculated $t_{LS}=2.01$. As anticipated from Eq.~(\ref{eq:tstat}) for $N=24$, the latter value is larger by a factor of precisely $\sqrt{23/22}$. Thus, the statistical significance of the LLS slope estimate, seemingly so far away from the world of random walks, can nonetheless be visualized as the ratio of two areas: that of a given DW to that of the standard deviation $\sigma_A$ for the ensemble of statistical realizations.

Characterizations of the LLS include maximum likelihood (when errors are normally distributed) and minimal variance, and figure prominently in data analysis \cite{papoulis1990probability,Bevington92,frieden2012probability,kay1998,press2007numerical}. By introducing the concept of a data walk (bridge), we have added another such result. There is an appealing complementarity about this result: from the random walk perspective, detrending amounts to seeking a simple zero of a function, while from the LLS perspective one has quadratic minimization.

Although there is a compelling symmetry to the annulment of the signed area DW in the absence of signals as depicted in Fig.~\ref{fig:walks}(d), the result is much more robust and holds even when the noise distribution is skewed, mean is not zero, and tails are broad. The Mathematica code provided as Supplementary Material~\cite{SM} includes an option of noise distributed as a one-sided $\beta$-distribution with parameters $\alpha = 5$, $\beta =1$, which can be easily adapted for any other distribution.

The reported LLS link to area distributions of returning walks suggests new questions such as the distribution of the LLS slope.  One can now explore the literature on Brownian bridges for comparison to distributions under discrete time bridges. For example, for $N \gg 1 $, $A$ (and the LLS slope) is a Gaussian random variable by the central limit theorem, with standard deviation $\sigma_A$ given by Eq.~(\ref{eq:stdevSlope}). In the absence of signal, its probability density function is
\begin{equation}
P(A,N) = \sqrt{\frac{6}{\pi\,\sigma\, N^3}}\, \exp\left (-\frac{6 A^2}{\sigma^2\, N^3}\right )\, .
\label{PDFA}
\end{equation}
Eq.~(\ref{PDFA}) parallels the probability density function for a Brownian bridge~\cite{horne2007analyzing}, keeping only the leading term for $\sigma_T$ where $T$ denotes time:
\begin{equation}
P(A,T) = \sqrt{\frac{3}{\pi\, T^3}}\, \exp\left (-\frac{3 A^2}{T^3}\right )\, .
\end{equation}
Note that the latter continuous process is restricted to Gaussian noise while the former discrete bridge applies to any distribution of steps, whether continuous or discrete.  

When the data samples are not equidistant in $x$, our extensive numerical experiments show that the association of the slope with the scaled trend (normalized DW area) $A(y)/A(x)$, although no longer exact, remains close.  This extends even to randomly spaced $\{ x_k\}$ such as for a Poisson process with exponentially distributed waiting times between data samples. Extensions to multidimensional data also appear promising.

\section*{Acknowledgments}
This work was supported in part by the National Science Foundation (NSF grant AGS-2217182) and in part by grant NSF PHY-2309135 to the Kavli Institute for Theoretical Physics (KITP). S.K. acknowledges startup funds provided by New York University.

\appendix

\section{Details for deriving Eq.~(\ref{eq:lintrend})}
\label{appendixA}
To evaluate $A(y) = - \sum_{j=1}^{N} z_j$, we use definitions of $x_k$, $y_k$, and $z_j$ in Eqs.~(\ref{eq:xk}), (\ref{eq:yk}), and (\ref{eq:PRW}).  For a linear function of unit slope and in the absence of noise, $z_j$ becomes 
\begin{equation}
z_j = \sum_{k=1}^{j} (x_k - \bar{x})
\end{equation}
The sample average $\bar{x}$ from $x_k = (k-1)/(N-1)$ can be evaluated explicitly:
\begin{align}
\bar{x} &= \frac{1}{N} \sum_{k=1}^{N} \frac{k-1}{N-1} = \frac{1}{N(N-1)} \sum_{k=1}^{N} (k-1) \nonumber \\ 
&= \frac{1}{N(N-1)} \left( \sum_{k=1}^{N} k - N \right)
\end{align}
where $\sum_{k=1}^{N} k$ is the $N$th partial sum of the natural numbers, also known as the triangular number.  In passing we note that the triangular number is also a binomial coefficient $\binom{N+1}{2}$. We obtain 
\begin{align}
\bar{x} &= \frac{1}{N(N-1)} \left( \sum_{k=1}^{N} k - N \right) = \frac{1}{N(N-1)} \! \left( \frac{1}{2} N (N+1) - N \right) \nonumber \\ 
&= \frac{1}{N(N-1)} \frac{N (N-1)}{2} = \frac{1}{2} \, .
\end{align}
Inserting this into the expression for $z_j$ and using the definition of $x_k$ yields
\begin{align}
z_j &= \sum_{k=1}^{j} \left(\frac{k-1}{N-1} - \frac{1}{2} \right) = \frac{1}{N-1} \sum_{k=1}^{j} (k-1) - \frac{j}{2} \nonumber \\
&=  \frac{1}{N-1}\left( \sum_{k=1}^{j} k \right) - \frac{j}{N-1} -\frac{j}{2} 
\end{align}
where again we use the triangular number to evaluate the remaining sum.  We obtain:
\begin{align}
z_j &=  \frac{1}{N-1}\left( \frac{1}{2} j(j+1) \right) - \frac{j}{N-1} -\frac{j}{2} = \frac{1}{N-1}\left( \frac{1}{2} j(j-1) \right) -\frac{j}{2} \nonumber \\ 
&= \frac{1}{N-1}\left( \frac{j^2-j}{2} \right) -\frac{j}{2} \, .
\label{eq:A5}
\end{align}
We can now sum the $z_j$s to evaluate $A(x)$:
\begin{align}
A(x) &= - \sum_{j=1}^{N} \left( \frac{1}{N-1}\left( \frac{j^2-j}{2} \right) -\frac{j}{2}  \right) \nonumber \\
&= - \frac{1}{2(N-1)} \left( \sum_{j=1}^{N} j^2 - \sum_{j=1}^{N} j \right) + \frac{1}{2} \sum_{j=1}^{N} j
\end{align}
and use again the partial sum results of Eq.~(\ref{eq:triangularnumber}). This yields
\begin{align}
A(x) &= - \frac{1}{2(N-1)} \left( \frac{N(N+1)(2N+1)}{6} - \frac{N(N+1)}{2} \right) + \frac{1}{2} \frac{N(N+1)}{2} \nonumber \\
&= -\frac{N(N+1)(2N+1)}{12(N-1)} + \frac{3N(N+1)}{12(N-1)} + \frac{3N(N+1)(N-1)}{12(N-1)} \nonumber \\
&= \frac{ N (N+1) (N - 1)}{12(N-1)} = \frac{N(N+1)}{12} \, .
\end{align}

\section{Details for deriving Eq.~(\ref{eq:reduce})}
\label{appendixB}

We begin with the left-hand side of Eq.~(\ref{eq:lsq}): 
\begin{equation}
\sum_{k=1}^N\, \sum_{j=1}^k \left ( y_j - \bar{y} \right )  =  (N-1)\, \sum_{k=1}^{N} y_k \,\left  ( \frac{1}{2}  - \frac{k-1}{N-1} \right )\, .
\end{equation} 
\vspace{-1mm} The first term of the sum on the left, i.e. $\sum_{k=1}^{N} \sum_{j=1}^{k} y_j$, gives $\sum_{k=1}^N \, (N-k+1)\, y_k$.  To see this, expand the sum as follows:
\begin{align}
  \sum_{k=1}^{N} \sum_{j=1}^{k} y_j =& \underbrace{y_1}_{k=1 \text{ term} } + \underbrace{(y_1 + y_2)}_{k=2 \text{ term} } + \underbrace{(y_1 + y_2 + y_3)}_{k=3 \text{ term} } \nonumber \\
  &+ \underbrace{(y_1 + y_2 + y_3 + y_4)}_{k=4 \text{ term}} + \cdots + \underbrace{(y_1 + y_2 + \cdots + y_N)}_{k=N \text{ term}} \, .
\end{align}
Upon enumerating terms, we observe that $y_1$ appears $N = N-1+1$ times, $y_2$ appears $N-1 = N-2+1$ times, $y_3 = $ appears $N-2 = N-3+1$ times, and so forth, until we reach the last term $y_N$, which appears only once.  Thus $y_k$ appears in the sum $N-k+1$ times, yielding 
\begin{equation}
    \sum_{k=1}^{N} \sum_{j=1}^{k} y_j = \sum_{k=1}^N \, (N-k+1)\, y_k \, .
\end{equation}
Adding this to Eq.~(\ref{eq:meanysum}) of the main text then yields Eq.~(\ref{eq:reduce}).

\end{document}